# Study of High Temperature Thermal Behavior of Alkyl and Perfluoroalkylsilane Molecules Self-Assembled on Titanium Oxide Nanoparticles

P. Prakash[a], U. Satheesh[a] and  D. Devaprakasam[a]*

[a]*NEMS/MEMS/Nanolithography Lab, Department of Nanosciences and Technology,*
*School of Nanosciences and Technology, Karunya University,*
*Coimbatore 641114, India.*

**Abstract**

We have studied high temperature thermal behavior of 1H, 1H, 2H, 2H-Perfluorooctyl-trichlorosilane (FOTS) and Octyltrichlorosilane (OTS) molecules self assembled on  titanium dioxide ($TiO_2$) nanoparticles using advanced microscopy and spectroscopy technique such as Scanning Electron Microscope (SEM), Dynamic Light Scattering (DLS), X-Ray Diffraction  (XRD) and Fourier Transform Infrared  Spectroscopy (FTIR). FOTS SAM and OTS SAM coated $TiO_2$ nanoparticles were heated to different temperature range from room temperature (RT) to 550 $^oC$. We characterized nano-microstructure and size distribution of FOTS SAM and OTS SAM coated $TiO_2$ nanoparticles, which were heat-treated to different peak temperatures, using SEM and DLS techniques. The thermal stability and degradation of FOTS SAM and OTS SAM coated $TiO_2$ were carried out using FTIR spectroscopy.  We found that FOTS SAM on $TiO_2$ is very stable up to 450 $^oC$ and OTS on $TiO_2$ is stable up to 250 $^oC$.  Peak frequency, peak intensity and full width half maxima (FWHM) of symmetric and asymmetric $CF_2$ and $CH_2$ confirms our observation. In this study, we successfully synergized the surface and temperature sensitive characteristics of FOTS/OTS SAM and $TiO_2$ nanoparticles in order to use them for highly demanding surface and temperature sensitive nanotechnology applications

## 1. INTRODUCTION

Self Assembled (SAM) and Langmuir-Blodgett (LB) Monolayer of organic molecules are widely used in various discipline of surface engineering to modify surface energy, adhesion, friction, surface electrical and electronics properties [1]. Recent advances in microscopy and spectroscopy techniques enabled us to use SAM and LB monomolecular films for variety of applications in the area of nanolithography, [2] nanoelectronics, [3] nanophotonics [4] and sensors [5, 6]. SAM can be easily patterned in two dimensional structures on metal, semiconductors and oxide surfaces by different methods like vapor phase [7-11],  liquid phase, [12] photolithography, [13] plasma lithography, [14] electrochemical deposition, [15-17] electron beam lithography, [18] x-ray lithography, [18] and atomic beam lithography. [18] Alkylsilane and perfluoroalkylsilane in iso-octane solution readily form covalent bond with hydrolyzed metal oxide surfaces. Closely packed alkylsilanes are easily getting polymerized in 2D planar direction by the formation of siloxane network by silanol groups [1, 12]. However in perfluoroalkylsilane the polymerization of silanol groups is limited by its size but perfluoroalkylsilane SAMs (FOTS, FDTS, OTS, FOTES, FOMMS and FOMDS) are highly hydrophobic, anti-friction, anti-stiction and have high thermal stability. [7, 8, 12, 19] FOTS SAM formed on different oxide surfaces have applications in anti-stiction coating on MEMS/ NEMS devices, [7, 11] organic thin film transistor (OTFT), [20] cell adhesion, protein adsorption, [21] wet chemical etching, [22] nano-micro-macrotribology, [23] and bio sensor. [24] $TiO_2$ nanoparticles have attractive potential applications like photo degradation, [25-27] corrosion studies, [28] self cleaning application, [29] solar energy, [30] electrochromic devices, [31] mechanical, [32] and biological area. [33]

We noted that both $TiO_2$ nanoparticles and FOTS/OTS SAMs have many surface and temperature sensitive properties and characteristics which were used in number potential applications in high end engineering and technological fields. Synergizing the properties of FOTS/OTS SAMs and $TiO_2$ nanoparticles, we can further enhance use of the combinatorial characteristics to potential new surface sensitive applications and also improve the thermal stability of the existing applications.

In this research work we have studied the thermo-energetic mechanism of self-assembled 1H,1H,2H,2H-perfluorooctyl trichlorosilane (FOTS) / Trichloro (Octadecyl) silane (OTS)  SAM coated $TiO_2$ nanoparticles by using Scanning Electron Microscopy (SEM), Fourier Transform Infrared Spectroscopy (FTIR) and  Dynamic Light Scattering (DLS) techniques and we attempted synergize the properties of $TiO_2$ nanoparticles and FOTS/OTS SAM in order to use them for highly demanding surface and temperature  sensitive nanotechnology applications.

---

\* Corresponding author. Tel.:     +91-422-2164488      fax: +91-422-614614
   *E-mail address*: devaprakasam@karunya.edu

## 2. Materials and Methods

*2.1 Materials details*

1H,1H,2H,2H-perfluorooctyl trichlorosilane (FOTS) (>97% Purity), Octyl trichlorosilane (OTS) (>97% Purity), Isooctane (>99.8% Purity) and $TiO_2$ nanoparticles were all purchased from Sigma Aldrich. The acetone (99% Purity) and methanol (99% Purity) were purchased from Merck.

*2.2 Preparation of SAM*

1mM FOTS and 1mM OTS in 25 ml isooctane were prepared and 1mg $TiO_2$ nanoparticles transferred to the freshly prepared FOTS/isooctane and OTS/isooctane solutions. Then the solution is ultrasonicated for 15 minutes to get homogeneous dispersion of coated nanoparticles and to minimize agglomeration. Using DLS technique, hydrodynamic particle size and distribution were obtained. Then FOTS SAM and OTS SAM coated $TiO_2$ nanoparticles were dried and preserved in vacuum desiccators. Then the samples were heat treated to various peak temperatures from RT to $550\,^{\circ}C$ for the FTIR and SEM analysis.

*2.3 Characterization techniques*

Nanostructure and surface morphologies of the FOTS/OTS SAM coated $TiO_2$ nanoparticles were observed by using JEOL model JSM-6390 scanning electron microscope. The characteristic vibration frequencies of FOTS SAM and OTS SAM on $TiO_2$ nanoparticles (before and after heat treatment) were studied by using SHIMADZU IR Prestige-21 model FTIR spectrometer. The particle size and distribution of the $TiO_2$ particle was analyzed using Malvern Particle (model) size analyzer. The crystalline phase of $TiO_2$ nanoparticles was characterized for X-Ray diffractometer (Shimadzu XRD 6000) instrument using Cu K ($\alpha$) 1.5406 Å wavelength.

## 3. Results and Discussions

*3.1 Nano-microstructure analysis*

Figure.1 (a) and (b) shows the schematic diagram of OTS SAM and FOTS SAM coated $TiO_2$ nanoparticles. 1Mm OTS and 1mM FOTS molecules in isooctane get hydrolyzed and trichlorosilane ($-SiCl_3$) molecules become silanol ($-Si(OH)_3$). Surface of $TiO_2$ get hydrolyzed for hydroxyl $-(OH)$ on the surface. A hydroxyl of silanol group and surface hydroxyl group of $TiO_2$ form Ti-O-Si covalent bond by chemical reaction [34, 35]. In the case of OTS SAM formation, lateral silanol group get polymerized for siloxane network. However in FOTS/SAM formation, the lateral polymerization is limited by its size [12]. This is clearly inferred from the FTIR spectra of FOTS SAM on $TiO_2$, where the -Si-O-Si- vibrational peaks at 1060 $cm^{-1}$ and 1070 $cm^{-1}$ are minimal compared to OTS SAM on $TiO_2$. Figure.2 (a) and (b) shows the SEM images of the FOTS annealed for different temperatures like (a) $150\,^{\circ}C$ for 1 hour and (b) $450\,^{\circ}C$ for 1 hour and Figure.2 (c) and (d) shows the SEM images of the OTS annealed for different temperatures like (c) $150\,^{\circ}C$ for 1 hour and (d) $450\,^{\circ}C$ for 1 hour. From the figure 2b and 2d it is very clearly seen that the surface morphology got changed by heating the both FOTS SAM and OTS SAM to $450\,^{\circ}C$. In the case of OTS SAM on $TiO_2$, it got desorbed from the $TiO_2$ surface at $450\,^{\circ}C$. However at the $450\,^{\circ}C$, in the case of FOTS SAM on $TiO_2$ the desorption is minimal compared to OTS SAM on $TiO_2$. When the temperature is further increased to above $450\,^{\circ}C$, the morphology got changed due to collapse of SAM on $TiO_2$ nanoparticles.

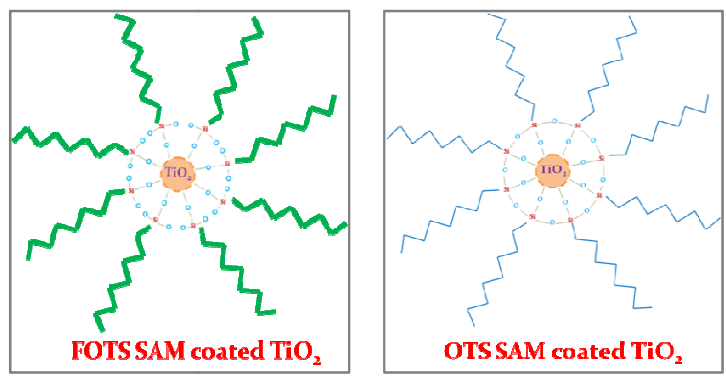

Fig.1. Shows (a) schematic diagram of FOTS SAM coated $TiO_2$ nanoparticle and (b) OTS SAM coated $TiO_2$ nanoparticle.

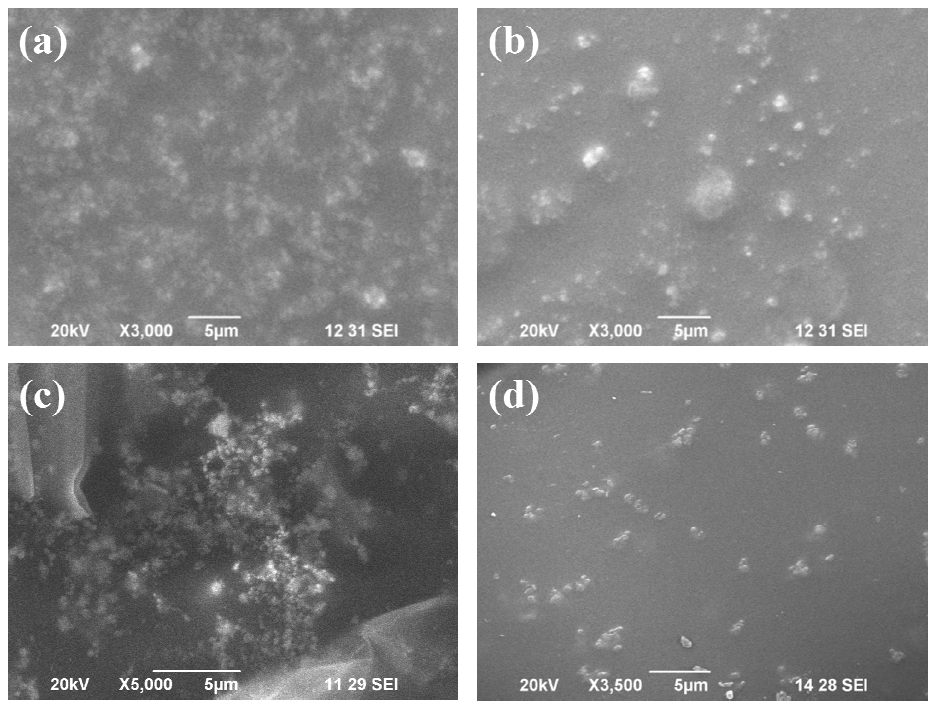

Fig.2. The nano-microstructure (a) and (b) FOTS SAM coated TiO$_2$ nanoparticles heat treated to 150°C and 450°C; (c) and (d) OTS SAM coated TiO$_2$ nanoparticles heat treated to 150°C and 450°C.

Figure.3 (a) shows XRD spectrum of TiO$_2$ nanoparticles. From the spectra it is observed that the TiO$_2$ nanoparticles are crystalline phase of anatase structure, which is confirmed by sharp peak around 2θ= 25°. Figure 3 (b) shows the particle size measurement for the TiO$_2$ in the isooctane solvent and FOTS/OTS SAM coated TiO$_2$ nanoparticles. The average particle size was found to be 255.3 nm for the TiO$_2$ in the solvent. When SAM material is added to the TiO$_2$ with solvent, the particle size started to increase. When OTS SAM material is added to TiO$_2$ coated nanoparticles, the hydrodynamic particle size was found to be 462.5 nm and when FOTS SAM material is added to TiO$_2$ coated nanoparticles, the particle size was found to be 310.3 nm. The increases in hydrodynamic radius of the TiO$_2$ particle indicate the FOTS and OTS molecules get adsorbed on the surface both physically and chemically. However it is noted that the high values of hydrodynamic radius associated with physical agglomeration of the TiO$_2$ nanoparticles.

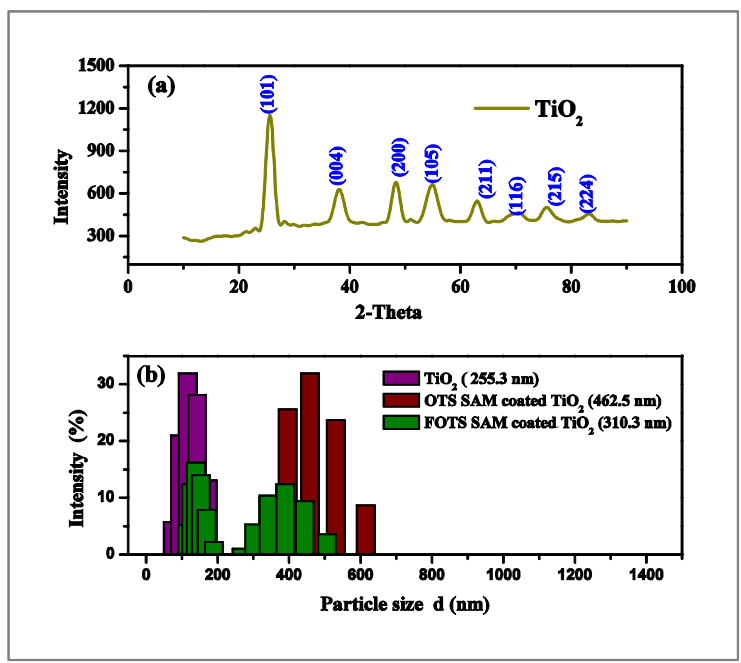

Fig.3. Shows (a) XRD spectrum of TiO$_2$ nanoparticles (b) The particle size distribution of uncoated, FOTS SAM and OTS SAM coated TiO$_2$ nanoparticles.

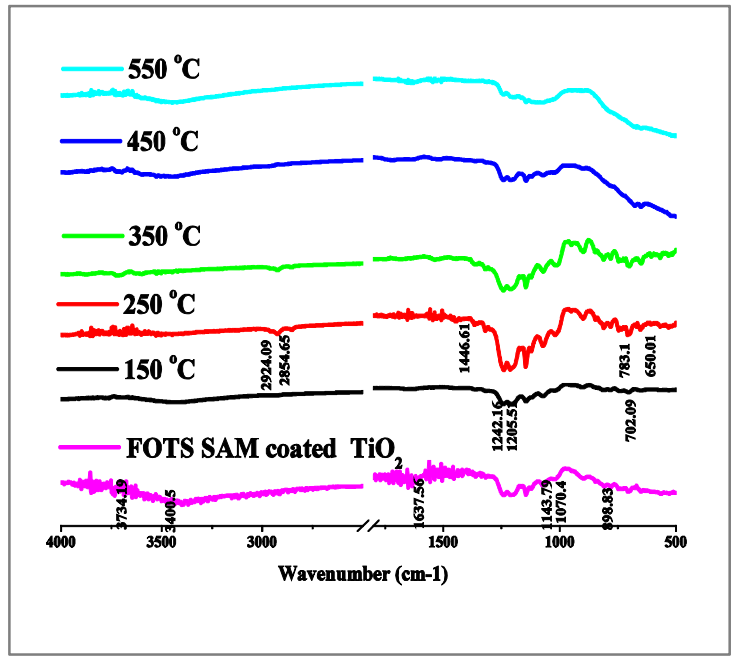

Fig.4. FTIR spectra show the thermal behavior of FOTS SAM coated TiO$_2$ nanoparticles heat treated from room temperature to 550°C.

*3.2 FTIR Analysis of FOTS SAM and OTS SAM on TiO$_2$ nanoparticles*

Figure.4 shows the FTIR spectrum of FOTS SAM coated TiO$_2$ nanoparticles, which were heat treated to different peak temperatures from room temperature (RT) to 550 °C. Figure.5 shows the symmetric and asymmetric stretching frequency of CF$_2$ functional group of FOTS SAM. Similarly Figure 6 and 7 shows FTIR spectrum of OTS SAM on TiO$_2$ and symmetric and asymmetric stretching vibrations of CH$_2$ which were heat treated to different peak temperatures from RT to 550 °C. Table 1 shows vibration frequency assignment of both FOTS and OTS molecules on TiO$_2$ nanoparticles. The peak frequency at 2850 cm$^{-1}$ and 2920 cm$^{-1}$ are due to symmetric and asymmetric stretching vibrations of CH$_2$ in alkyl chain. The peaks at 1143 cm$^{-1}$ and 1242 cm$^{-1}$ are the symmetric and asymmetric stretching frequency of CF$_2$ in the perfluoroalkyl chain. The peaks found around 1060 cm$^{-1}$ and 1070 cm$^{-1}$ attributed to polymerized siloxane network Si-O-Si. The anchoring bond Ti-O-Si peaks are observed at 885 cm$^{-1}$. Especially in FOTS there are un-reacted silanol groups Si-OH and their peaks observed around 898 cm$^{-1}$. Strong and weak frequency of C-C-C group vibrations are observed at 1122 cm$^{-1}$ and 1205 cm$^{-1}$. The peaks at 572 cm$^{-1}$ and 1018.41 cm$^{-1}$ are due to the C-C vibration.

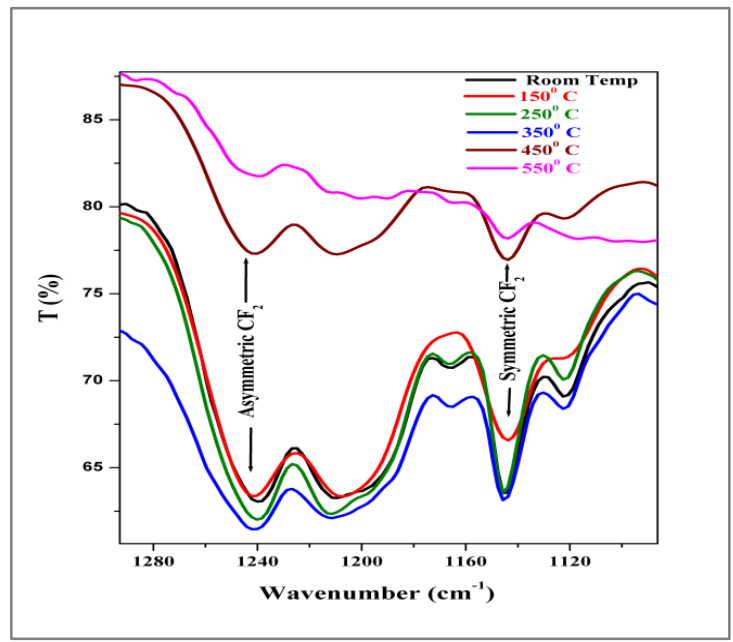

Fig.5. Shows symmetric and asymmetric stretching vibrations of CF$_2$ functional groups.

Table.1. Shows functional groups vibration in FOTS SAM/ OTS SAM coated TiO2 nanoparticle.

| Peak Frequency (cm$^{-1}$) | Assignment Normal | References |
|---|---|---|
| 455.2, 467,447 | Ti-O, Ti-O-Ti | 34, 36, 37, 38 |
| 50.01 | C-H bending vibration, C-H deformation | 39 |
| 688 | Ti-O-O band | 37 |
| 702.09 | $CF_2$ | 12 |
| 783.1, 794 | Si-O groups vibration | 40 |
| 898.83, 885 | Si-OH, Ti-O-Si | 12, 34,35 |
| 1070.49, 1060 | -Si-O-Si- | 12, 41, 42 |
| 1122 | C-C stretching (symmetric), C-C-C strong | 12 |
| 1143.79 | Symmetric $CF_2$ | 12, 40 |
| 1205.51 | C-C-C , C-C | 12 |
| 1242.16 | Asymmetric $CF_2$ | 12, 40 |
| 1446.61 | $CH_2$ in-plane deformation | 39, 40 |
| 1462 | Asymmetric vibration $CH_3$ | 26 |
| 1637.56, 1641.42 | Ti-OH , bending vibration of OH | 27 |
| 2854.65, 2856 | C-H stretching (symmetric) | 41 |
| 2924.09 , 2964 | C-H stretching (asymmetric) | 41 |
| 3400.5, 3363, 3462 | OH stretching , Ti-OH stretching | 27, 31,39 |

The C-H out of plane bending vibrations is found at 690 cm$^{-1}$ and 850 cm$^{-1}$. Absorption peaks found at 400 cm$^{-1}$ and 800 cm$^{-1}$ were correspond to Ti-O-Ti peak frequency. There various characteristics vibrations associated with FOTS SAM and OTS SAM molecules on TiO$_2$ nanoparticles are given the Table 1. From these studies, we observed that both FOTS and OTS molecules self-assembled on TiO$_2$, they were chemically adsorbed on the TiO$_2$ nanoparticles surface through Ti-O-Si bond. We also noted that silanol groups in OTS SAM laterally polymerized and forms 2D siloxane network, which were limited in the case of FOTS SAM due to its size.

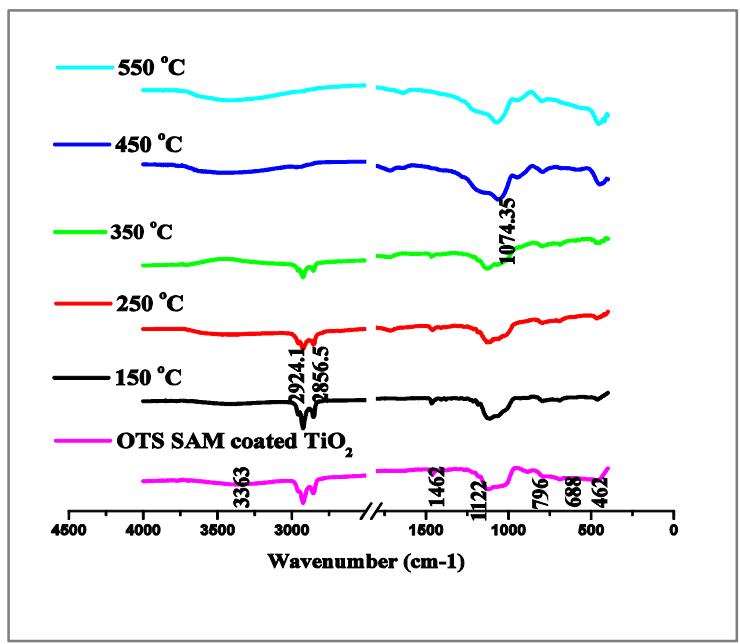

Fig.6. FTIR spectra of show thermal behavior for OTS SAM coated TiO$_2$ nanoparticles heat treated from room temperature to 550$^o$C.

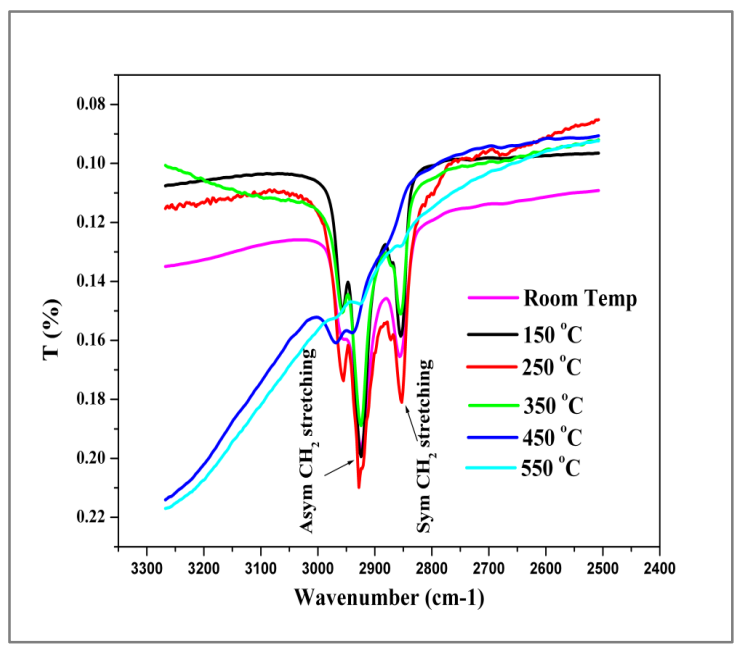

Fig.7. Shows symmetric and asymmetric stretching vibrations of CH$_2$ functional groups.

*3.3 Thermal stability of FOTS SAM and OTS SAM on TiO$_2$ nanoparticles*

Figure.8 (a)-(d) shows the peak frequency, integrated intensity and FWHM symmetric and asymmetric vibrations of CF$_2$ functional group of the FOTS SAM heat treated to different peak temperature from RT to 550 °C. It is observed that there are a priori conformational and orientational defects in the freshly prepared FOTS SAM. It is observed that when the temperature increased from the RT to 250 °C, the initial defects gets healed which is evident from increase in peak frequency, integrated intensity of CF$_2$ vibrations. Further heating started to introduce conformational defects in the molecular architecture of FOTS SAM molecules. However due to uncoiling and tilting effects of helical conformation, the orientational order increases which in turn shows increase in crystalline nature of the functional group and backbone chain which is evident from FWHM characteristics. Upon the further heat treatment between 250 °C to 450 °C the defects get accumulated in the molecular architecture of FOTS SAM and it becomes more disordered state, this is clearly seen in the decrease of peak frequency and integrated intensity. Beyond 450 °C until 550 °C the FOTS SAM collapses due saturation of defects and beyond 550 °C subsequent scission of Ti-O-Si bond result in desorption FOTS SAM from the TiO$_2$ surface.

Figure.9 (a)-(d), shows symmetric, asymmetric, FWHM of CH$_2$ stretching frequencies of OTS SAM on TiO$_2$ nanoparticles which were heat treated to different temperature from RT to 550 °C. When the temperature is increased, initially a priori defects in the system get healed but after the initial recovery the defects get introduced in the molecular architecture of OTS SAM molecule, decrease in peak frequency, integrated intensity and increase in FWHM, indicates both conformational and orientational defects continuously increases. Upto 350 °C the defects gets accumulated in the OTS SAM and the accumulated defects result in its collapse and beyond 350 °C, the scission of alkyl -Si bond result in desorption backbone chain, however presence of Si-O-Si indicates the 2D siloxane network very much intact and stay on the surface even beyond this temperature upto 550 °C.

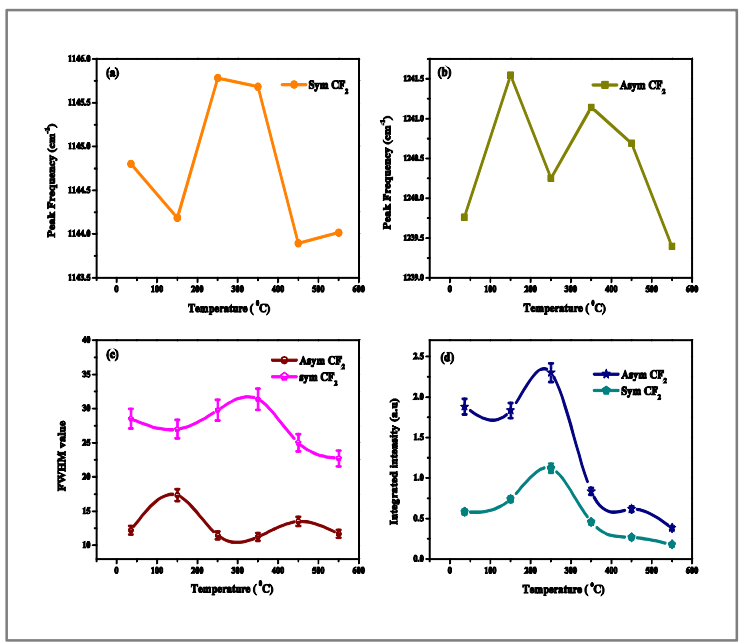

Fig.8. Shows thermal behavior of $CF_2$ symmetric and asymmetric vibrations (a) & (b) the peak frequency (c) FWHM and (d) Integrated Intensity.

From this analysis it can be inferred that both FOTS SAM and OTS SAM are suitable for functionalizing surface of the $TiO_2$ nanoparticles which could used high temperature applications. FOTS SAM is stable upto 450 °C and OTS SAM stable upto 250 °C on $TiO_2$ nanoparticles. Depends upon the requirement we can use either system or even combination of them for the potential applications.

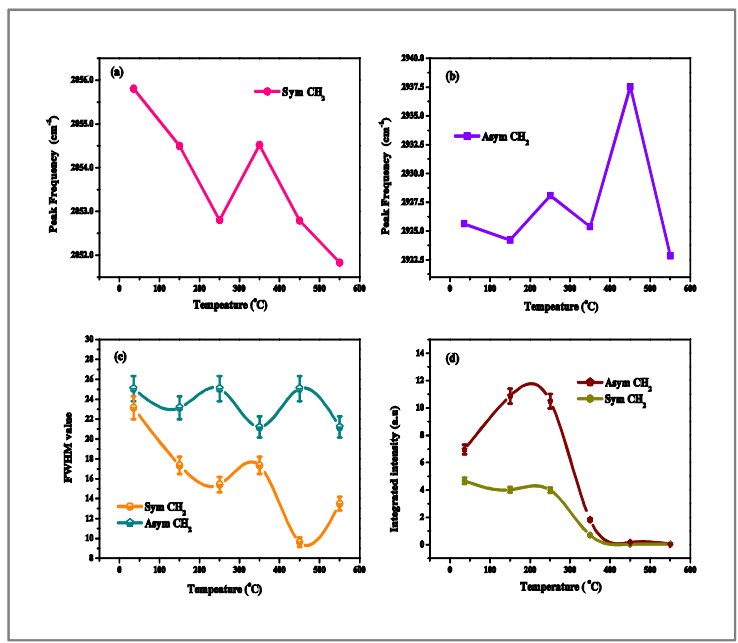

Fig.9. Shows thermal behavior of $CH_2$ symmetric and asymmetric vibrations (a) & (b) the peak frequency (c) FWHM and (d) Integrated Intensity.

## 4. CONCLUSION

The thermal behavior of FOTS and OTS molecules self-Assembled on $TiO_2$ nanoparticles were studied using advanced microscopy and spectroscopy techniques. When the OTS SAM on $TiO_2$ nanoparticles is subjected to heat treatment from RT to 550 °C, it undergoes various ordered to disordered state due to various conformational and orientational defects introduced into the molecular architecture. This is very evident from the FTIR spectra at various heat treatment temperature, peak frequency, and integrated intensity and FWHM of symmetric and asymmetric vibrations of $CH_2$ at 2850 $cm^{-1}$ and 2920 $cm^{-1}$. OTS SAM became disordered when it is heat treated to 250 $^{o}$C and beyond. However it has been observed that the polymerized siloxane networks (The -Si-O-Si- peaks at 1060 cm$^{-1}$ and 1070cm$^{-1}$.) are very stable on TiO$_2$ surface. In the case of FOTS SAM, FTIR spectra at various heat treatment temperature, peak frequency, integrated intensity and FWHM of CF$_2$ symmetric and asymmetric vibrations at 1144 cm$^{-1}$ for symmetric and 1245 cm$^{-1}$ shows that it is highly stable up to 450 $^{o}$C and beyond it become more disordered. At low temperature the freshly prepared FOTS SAM has number of *a priori* defects in, further heat treatment up to 250 $^{o}$C, it became more ordered state. Between 250 $^{o}$C and 450 $^{o}$C heat treatment, FOTS/SAM on TiO$_2$ undergoes various ordered to disordered states because of orientational and conformational defects. Moreover, lateral polymerization of silanol group is restricted in FOTS/SAM on TiO$_2$ due to its size caused by helical conformations. From these studies we conclude that both FOTS SAM and OTS SAM are suitable candidate for functionalizing TiO$_2$ nanoparticles which could be used for high temperature applications.

**Acknowledgments**

I thank the Dr. Paul Dinakaran, Chancellor, Karunya University, Coimbatore for providing the Silver Jubilee Fellowship to Mr. P. Prakash for PhD research work. We thank technical staff members of Centre for Research in Nanotechnology, School of Nanosciences and Technology, Karunya University for their technical help.